\documentclass[12pt]{iopart}

\expandafter\let\csname equation*\endcsname\relax
\expandafter\let\csname endequation*\endcsname\relax
\usepackage{amsmath}

\usepackage{graphicx}
\usepackage{rotating}
\usepackage{cite}
\usepackage{color}
\usepackage{soul}
\usepackage{siunitx}
\usepackage[colorlinks=true
,urlcolor=blue
,anchorcolor=blue
,citecolor=blue
,filecolor=blue
,linkcolor=blue
,menucolor=blue
,linktocpage=true
,pdfproducer=medialab
,pdfa=true
]{hyperref}
\usepackage{soul}
\usepackage[table]{xcolor}
\usepackage{multirow}
\usepackage{comment}

\begin{document}
	
	\setlength{\parindent}{0pt}
	
	\title[ ]{Frequency-dependent electron power absorption mode transitions in capacitively coupled argon-oxygen plasmas}
	
	\author{ A. Derzsi$^{1}$, M. Vass$^{1,2}$, R. Masheyeva$^{1,3}$, B. Horv\'ath$^1$,  Z. Donk\'o$^1$, P. Hartmann$^1$}
	
	\address{
		$^1$ Institute for Solid State Physics and Optics, Wigner Research Centre for Physics, 1121 Budapest, Hungary\\
		$^2$ Department of Electrical Engineering and Information Science, Ruhr University Bochum, 44780 Bochum, Germany\\
        $^3$ Al-Farabi Kazakh National University, Institute of Experimental and Theoretical Physics, 050040 Almaty, Kazakhstan
	}
	
	\ead{derzsi.aranka@wigner.hu}
	
	\begin{abstract}
		Phase Resolved Optical Emission Spectroscopy (PROES) measurements combined with 1d3v Particle-in-Cell/Monte Carlo Collision (PIC/MCC) simulations are performed to investigate the excitation dynamics in low-pressure capacitively coupled plasmas (CCPs) in argon-oxygen mixtures. The system used for this study is a geometrically symmetric CCP reactor operated in a fixed mixture gas composition, at fixed pressure and voltage amplitude, with a wide range of driving RF frequencies (2~MHz~$\le f \le$~15~MHz). The measured and calculated spatio-temporal distributions of the electron impact excitation rates from the Ar ground state to the Ar~$\rm{2p_1}$ state (with a wavelength of 750.4~nm) show good qualitative agreement. The distributions show significant frequency dependence, which is generally considered to be predictive of transitions in the dominant discharge operating mode. Three frequency ranges can be distinguished, showing distinctly different excitation characteristics: (i) in the low frequency range ($f \le$~3~MHz), excitation is strong at the sheaths and weak in the bulk region; (ii) at intermediate frequencies (3.5~MHz~$\le f \le$~5~MHz), the excitation rate in the bulk region is enhanced and shows striation formation; (iii) above 6~MHz, excitation in the bulk gradually decreases with increasing frequency. Boltzmann term analysis was performed to quantify the frequency dependent contributions of the Ohmic and ambipolar terms to the electron power absorption. 
	\end{abstract}
	
\section{Introduction} \label{sec:Introduction}

Capacitively coupled plasmas (CCPs) driven by radio frequency (RF) waveforms have a wide range of applications in the semiconductor industry and are basic tools in biomedical applications \cite{Liebermann_book,Makabe_book,Chabert_book,Makabe08,Makabe_2019}. Study of processing plasmas is driven by the motive of understanding the complex physical and chemical interactions as well as improving performance and control in such systems \cite{Bogdanova_2f_2022, Wu_2023, Liu_2022_fundamental}. In plasma processing applications electronegative gases are frequently diluted with electropositive gases. Argon mixed with reactive gases is particularly important as it can enhance the etching process \cite{Takechi-ArO2_2001}. 

Gas discharges in argon-oxygen mixtures are often used for sputtering deposition, to grow SiO$_2$ dielectric films on silicon, to etch photoresist and polymer films, as well as for sterilization of medical instruments and surface activation \cite{Yonemura-ArO2_magnetron_2003,Makabe-ArO2_2006,Xu-ArO2_2007,Kutasi-ArO2_2010,Izdebska_2023}. 
Over the years, the plasma properties of low-pressure argon-oxygen mixture CCPs have been studied extensively via modeling and numerical simulations as well as experimentally \cite{McMillin-ArO2_1996,Benyoucef-ArO2_2011,Yamamoto-ArO2_2019,Jiang-2f-ArO2_2019,Oberberg-Ar-O2_2020,Nikolic-ArO2_2021,Booth_2009,Bai_ArO2_global_2023}.
A reaction set for modeling argon-oxygen CCPs was introduced  in \cite{babaeva2005oxygen} and the role of ionization, resonant and nonresonant charge-exchange collisions in the formation of the ion energy distribution at the electrodes was studied by Particle-in-Cell/Monte Carlo Collisions (PIC/MCC) simulations.
PIC/MCC and fluid simulations of argon-oxygen CCPs were performed in \cite{lee2006particle} to study the effects of the gas pressure and the ratio of argon and oxygen in the mixture on the plasma density, space potential, electron temperature and ion energy distribution. A global (volume averaged) model for argon-oxygen discharges was proposed in \cite{gudmundsson2007oxygen} to determine which reactions are important in the discharge model. An extensive set of processes was also provided in \cite{Bogaerts-ArO2-hybrid-2009} in the frame of a hybrid Monte Carlo-fluid model. 
The mechanisms of Ar metastable generation was investigated in argon-oxygen CCPs by two-dimensional computer simulations in \cite{Shahid_Kushner_1997}, showing that small amount of oxygen in the mixture decreases the Ar metastable atom density due to quenching by O and O$_2$ and change their spatial density profile due to a transition to an electronegative plasma. 
In dual-frequency argon-oxygen CCPs, the effects of the external discharge parameters, namely the effects of the frequency and power of the low frequency source and the gas pressure on the energy distributions of ions bombarding the electrodes, were studied both experimentally and by simulations \cite{Liu-ArO2_2013}.
Recently, the use of tailored voltage waveforms in geometrically asymmetric CCPs sustained in argon-oxygen mixtures was investigated computationally with the goal of shaping the energy and angular distributions of electrons incident onto the substrate to address positive surface charging inside nanometer scale high-aspect-ratio features \cite{Kruger_2021}. 

The characterization of CCPs in terms of operation modes is linked to the electron power absorption in such systems. In low-pressure CCPs, various discharge operation modes can be observed. The most common ones are the $\alpha$-mode and the $\gamma$-mode \cite{Belenguer1990} in electropositive gases, while the drift-ambipolar (DA) mode \cite{Schulze2011} and the striation (STR) mode \cite{Liu2016, Liu2017, 
Wang_2019, Skarphedinsson_2020, Proto_2021} are characteristic of electronegative gases. 
In the $\alpha$-mode, the ionization is caused by energetic electrons accelerated in the vicinity of the edges of the expanding sheaths, resulting in ionization peaks ($\alpha$-peaks) at the expanding sheath edges. In the $\gamma$-mode, the ionization peaks within the sheaths ($\gamma$-peaks) and the ionization is primarily caused by secondary electrons emitted from the electrodes accelerated in the sheaths. In the DA-mode, the ionization is mainly concentrated in the central bulk region and at the collapsing sheath edges. Here, the ionization is generated by electrons accelerated by the drift electric field in the bulk (caused by the low conductivity of the plasma bulk), and by the ambipolar electric fields at the sheath edges (caused by the strong electron density gradients). In the STR-mode (which develops when both positive and negative ions can react to the fast variation of the RF electric field), the ionization, concentrated within the bulk, exhibits features called ``striations''. These structures are due to the modulation of the electric field and the electrons' power absorption in the bulk. 

By varying the external control parameters, transitions between the different operation modes of low-pressure CCPs can be observed. In pure oxygen CCPs, by changing the gas pressure a transition between the $\alpha$-mode and the DA-mode was found \cite{ Derzsi_2017,Gudmundsson_2019}. Similar mode transitions were found to be induced by changing the gap distance \cite{Gudmundsson_2019, Hyo-Chang_2019_O2}, the driving frequency \cite{Derzsi_2017,Gudmundsson_2018}, the driving voltage waveform \cite{Derzsi_2015, Derzsi_2017, Gibson_2017, Gudmundsson_2017, Donko_2017_PPCF, Donko_2018}, as well as the external magnetic field \cite{Wang_2020}. By increasing the driving frequency at a constant pressure \cite{Gudmundsson_2018}, as well as by increasing the pressure or the electrode gap \cite{Gudmundsson_2019}, a transition from a hybrid DA-$\alpha$ mode to a pure $\alpha$-mode was observed. A transition from the STR-mode to $\gamma$-mode due to enhanced secondary electron emission was also found in oxygen CCPs \cite{Wang_2019}, as well as a transition between the $\alpha$-mode and $\gamma$-mode \cite{Lisovskiy-O2-2004}.
In argon CCPs, $\alpha$-$\gamma$ mode transitions were found under a wide range of discharge conditions \cite{Belenguer1990, Donko_2010, Schulze_2011}.

Here, we study the electron power absorption and excitation dynamics in CCPs operated in mixtures of 70\%~argon and 30\%~oxygen (volumetric ratio). Phase Resolved Optical Emission Spectroscopy (PROES) measurements combined with PIC/MCC simulations are performed in a geometrically symmetric CCP reactor in a wide frequency range (2~MHz~$\le f \le$~15~MHz) at constant pressure (120~Pa) and peak-to-peak voltage (350~V).
PROES \cite{Gans_2004, Gans_2010, Schulze_JPD_2010} is considered to be an experimental tool which can reveal the electron power absorption and discharge operation mode in CCPs. PROES images are frequently used to formulate statements on the operation mode of low-pressure CCP discharges despite the fact that PROES provides information on the spatio-temporal distribution of the electron-impact excitation dynamics from the ground state into the selected excited atomic state in the discharge, while the discharge operation mode is determined by the spatio-temporal distribution of the ionization dynamics. Recently, the applicability of PROES to probe the discharge operation mode was tested in low-pressure CCPs in pure neon \cite{Horvath_2020} and in neon-oxygen mixtures \cite{Derzsi_2022_NeO2}, and limiations were revealed especially around the transition regime between the $\alpha$-mode and the $\gamma$-mode.

The computational investigation of the spatio-temporal dynamics of the electron power absorption is based on the Boltzmann term analysis, a computational diagnostic tool, which is capable of providing a complete description of the electron power absorption in CCPs \cite{Wilczek_2020,Schulze2018_Boltzmann}. This method, first proposed in \cite{surendra1993moment} and later revisited in \cite{lafleur2014electron}, has recently been applied to CCPs under various conditions: in inert gases \cite{Schulze2018_Boltzmann,Wilczek_2020,Vass_2020ar,wang2022validity,VMBulk,liu2022fundamental,vass2022frequency,wu2022note}, in electronegative gases \cite{Vass_2020,Proto_2020,Proto_2021,Derzsi_2022_NeO2}, in CCPs at atmospheric pressure \cite{VMCOST}, and in magnetized CCPs as well \cite{bocong1,bocong2,eremin2023electron,Wang_2020}.

The paper is structured in the following way. In section~\ref{sec:Methods}, the experimental setup, the simulation method and the argon-oxygen discharge model is described, as well as the the Boltzmann term method. The results are discussed in section~\ref{sec:Results}. The conclusions are drawn in section~\ref{sec:Conclusions}.

\section{Methods and discharge conditions} \label{sec:Methods}

\subsection{Experimental method} \label{sec:Experimental}
A geometrically symmetric plasma reactor (our ``Budapest v.3'' cell) is used for the measurements \cite{Derzsi_2022_NeO2}. In the discharge cell, the plane parallel electrodes, made of stainless steel, with identical diameters of 14~cm, are situated within a quartz cylinder. One electrode is driven by a RF voltage, while the other electrode is grounded. The distance between the electrodes is set to $L=2.5$~cm. The background gas is an argon-oxygen mixture (70\%~Ar--30\%~O$_2$). The gas pressure is kept constant at $p=120$~Pa. The driving frequency, $f$, is varied between 2~MHz and 15~MHz at a constant peak-to-peak voltage of $V_{\rm{pp}} = 350$~V.
PROES measurements \cite{Gans_2004, Gans_2010, Schulze_JPD_2010} are performed by using a fast-gateable ICCD camera (4 QuickE, Stanford Computer Optics). The optical emission from the 
Ar~$\rm{^2P_1}$ excited state (also denoted as 
$\rm{2p_1}$ in the simplified Paschen notation)  with a wavelength of 750.4~nm is measured (by applying an interference filter with a central wavelength of 750~nm and a spectral full width at half maximum of $\sim$10~nm), 
from which the electron impact excitation rate from the ground state into the observed state is calculated as introduced in \cite{Schulze_JPD_2010}. A more detailed description of the experimental setup and diagnostics is given in \cite{Derzsi_2022_NeO2}. 

\subsection{Simulation method} \label{sec:Simulation}
The simulations are based on a one dimensional in space and three dimensional in velocity space (1d3v) Particle-in-Cell/Monte Carlo Collisions (PIC/MCC) simulation code \cite{Birdsall_Book,Verboncoeur2005,Schneider,Donko_2011_PSST,Sun_2016,Radmilovic-Radjenovic2009}. The particles traced in the simulations of the argon-oxygen gas mixture are electrons, Ar$^+$ ions, fast Ar atoms (Ar$^{\rm f}$), ${\rm O_2^+}$ ions, ${\rm O^-}$ ions and fast ${\rm O_2}$ molecules (O$_2^{\rm f}$). $\rm{O_{2}(a^{1}\Delta_{g})}$ metastable molecules are also considered in the model as continuum species. In total, 61 collision processes are included in the argon-oxygen discharge model. 

\begin{figure}[ht]
	\centering
	\includegraphics[width=.5\linewidth]{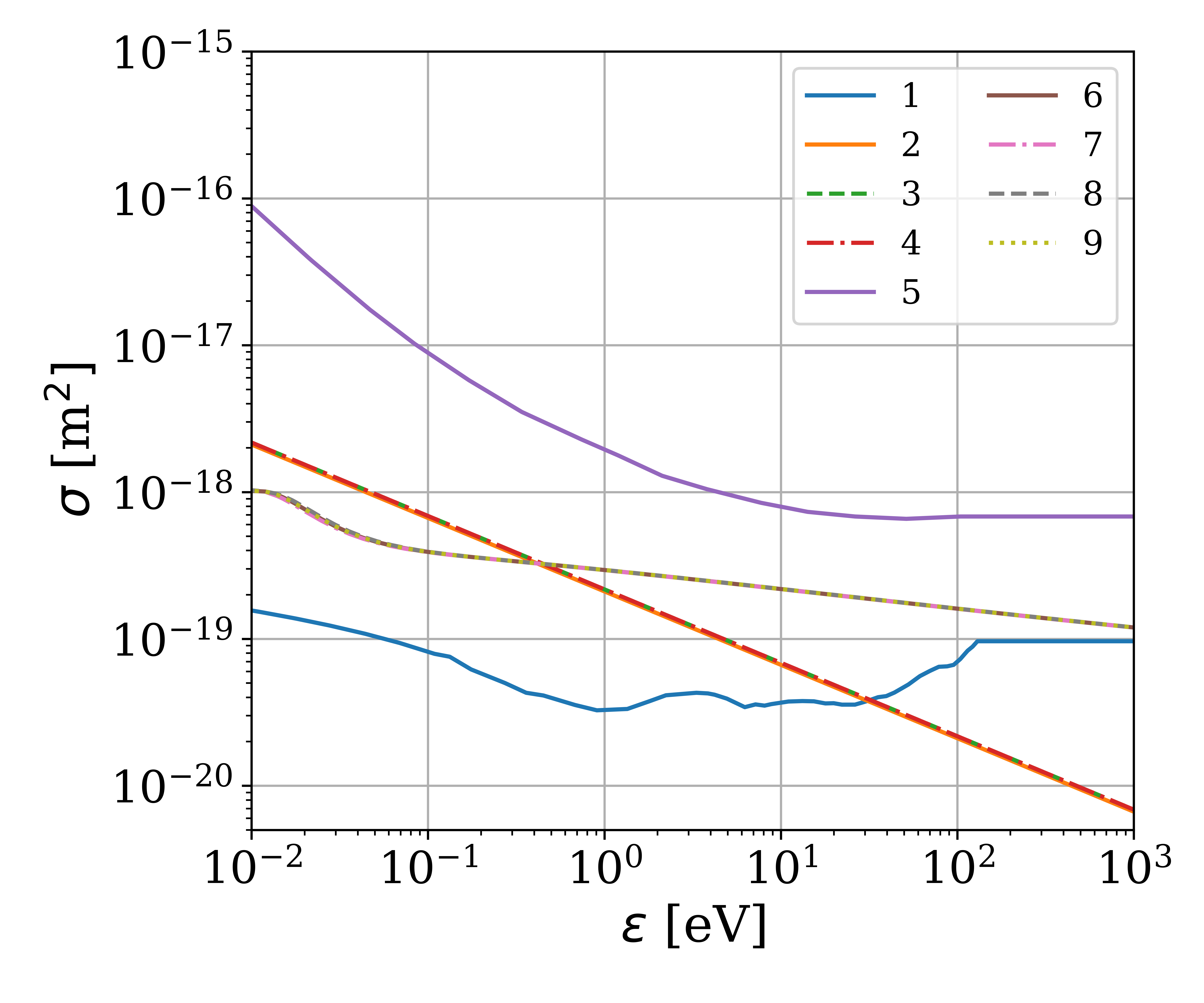}
	\caption{Cross sections of the collision processes listed in table~\ref{table:Argon-Oxygen-mixture} (processes 1--9), as a function of the kinetic energy (considered in the center-of-mass frame). ``Cross processes'' between oxygen and argon species:
	charge transfer and isotropic elastic scattering between Ar$^+$ ions and O$_2$ molecules (processes 1 and 2, respectively), isotropic elastic scattering between $\rm{O_2^+}$ ions and Ar atoms (process 3), and $\rm{O^-}$ ions and Ar atoms (process 4), and mutual neutralisation of $\rm{O^-}$ with Ar$^+$ (process 5). Cross sections of the collision processes for fast Ar atoms (Ar$^{\rm f}$) and fast O$_2$ molecules (O$_2^{\rm f}$) (processes 6--9): elastic scattering between Ar$^{\rm f}$ and Ar atoms/O$_2$ molecules (process 6/8) and elastic scattering between O$_2^{\rm f}$ and O$_2$ molecules/Ar atoms (process 7/9). Note that the cross sections overlap in case of processes 2--4 and processes 6--9.}
	\label{fig:argon-oxygen_collisions}
\end{figure}

For $\rm{e^{-} + Ar}$ collisions, the Hayashi cross section set (comprising 27 collision processes, including 25 excitation channels) \cite{Hayashi} is used (available at the LxCat database \cite{LXCat_2012,LXCat_2017,LXCat_2021}). For $\rm{Ar^{+} + Ar}$ collisions (including isotropic and backward elastic scattering processes) the cross section data from Phelps \cite{phelps1994application} are used. The cross sections adopted for the collisions of electrons with O$_2$ molecules, O$_2^+$ ions and O$^-$ ions, as well as for the collisions of oxygen species with these targets and $\rm{O_{2}(a^{1}\Delta_{g})}$ metastable molecules (23 collision processes in total) are the same as those introduced in \cite{Derzsi_2015} and used previously in simulation studies of CCPs operated in pure oxygen \cite{Derzsi_2015,Derzsi_2017,Donko_2018,Hyo-Chang_2019_O2,Vass_2020, Vass_2021} and neon-oxygen mixtures \cite{Derzsi_2022_NeO2}.
More details as well as plots of these cross sections can be found in the papers quoted above and these are not repeated here. The above collision processes are complemented with ``cross processes'' between oxygen and argon species (5 processes) and collision processes for fast neutrals (4 processes), shown in table~\ref{table:Argon-Oxygen-mixture}. The cross sections of these processes are plotted in figure~\ref{fig:argon-oxygen_collisions}.

\begin{table}[ht]
\caption{\label{n2} The list of ``cross processes'' between oxygen and argon species and collision processes for fast neutrals (Ar$^{\rm f}$ and O$^{\rm f}_{2}$). The cross sections of processes 6--9 are calculated as described in \cite{Derzsi_2022_NeO2}.}
\footnotesize
\begin{tabular}{@{}llll}
\br
\#&Reaction&Process&References\\
\mr
1& Ar$^+$ + O$_2$ $\longrightarrow$ Ar + O$_2^+$ &  Charge transfer  & \cite{Flesch_1990}\\ 
2& Ar$^+$ + O$_2$ $\longrightarrow$ Ar$^+$ + O$_2$ &  Isotropic elastic scattering  & \cite{Langevin}\\ 

3& O$_2^+$ + Ar $\longrightarrow$ O$_2^+$ + Ar &  Isotropic elastic scattering  & \cite{Langevin}\\ 

4& O$^-$ + Ar $\longrightarrow$ O$^-$ + Ar &  Isotropic elastic scattering  & \cite{Langevin}\\   

5& O$^-$ + Ar$^+$ $\longrightarrow$ O + Ar &  Mutual neutralisation  & \cite{gudmundsson2013benchmark} \\  
\mr

6&$ \rm{Ar^{f} + Ar \longrightarrow Ar^{f} + Ar} $  &  Isotropic elastic scattering & 
\\ 

7&$ \rm{O^{f}_{2}+O_{2}}  \longrightarrow  O^{f}_{2}+O_{2} $  &  Isotropic elastic scattering &  \\ 

8&$ \rm{Ar^{f} + O_{2} \longrightarrow Ar^{f} + O_{2}} $  &   Isotropic elastic scattering &  \\ 

9&$ \rm{O^{f}_{2}+Ar  \longrightarrow  O^{f}_{2}+Ar }$  &  Isotropic elastic scattering &  \\ 
\br
\end{tabular}\\
\label{table:Argon-Oxygen-mixture}
\end{table}

The collision processes listed in table~\ref{table:Argon-Oxygen-mixture}, which are specific of the argon-oxygen discharge model are discussed below. Process 1 is a nonresonant charge transfer reaction caused by two different atomic states of the projectile: $\rm{Ar^+(^2P_{3/2})}$ and $\rm{Ar^+(^2P_{1/2})}$. They have different charge exchange cross sections within the same order of magnitude and with similar trends according to Flesch \emph{et al.} \cite{Flesch_1990}. As the triplet and the singlet states are created with ratios 2:1 in ionization, a weighted average of the two cross sections is used, correspondingly.
Processes 2, 3, and 4 (isotropic elastic scattering between Ar$^+$ ions and O$_2$ molecules, between O$_2^+$ ions and Ar atoms, and between O$^-$ ions and Ar atoms, respectively) are treated with their Langevin cross section: $\sigma_{\rm L} = \sqrt{\frac{\alpha^\ast \pi e^2}{\epsilon_0 \mu}}\frac{1}{g}$,
where $\alpha^\ast$ is the polarizibility of the target, $\mu$ is the reduced mass, and $g$ is the relative velocity. The polarizibility values are: $\alpha^\ast$(O$_2$)~=~1.562~$\times~10^{-30}$~m$^{3}$ and $\alpha^\ast$(Ar)~= 1.664 $\times 10^{-30}$ m$^{3}$ \cite{Polarizabilites}. The cross sections of processes 2, 3, and 4 are nearly equal to each other. Due to the lack of data, process 5 (mutual neutralisation of O$^-$ with Ar$^+$ ion) is treated with the same cross section as the mutual neutralisation between O$^-$ and O$_2^+$ \cite{gudmundsson2007oxygen}.
 The cross sections for fast neutral collisions in argon--oxygen mixtures (processes 6--9) are calculated based on the pair-potential between the particles, for which the Lennard-Jones type is assumed (similarly to the case of fast neutrals in neon--oxygen mixtures, discussed in details in \cite{Derzsi_2022_NeO2}). The cross sections of these processes are also nearly equal to each other.

The heating of the background gas due to elastic collisions of fast neutrals and ions with thermal atoms/molecules of the background gas and heating up of the electrodes due to inelastic collisions of plasma particles with the electrodes is taken into account in the discharge model (see details in \cite{Derzsi_2022_NeO2}). 
As surface processes, secondary electron emission due to Ar$^+$ and O$_2^+$ ions, elastic reflection of electrons and surface quenching of $\rm{O_{2}(a^{1}\Delta_{g})}$ metastable molecules are considered in the model by constant surface coefficients. The elastic electron reflection coefficient is set to $\eta_{\rm{e}}=0.7$ \cite{Schulenberg21}. The secondary electron emission coefficient is set to 0.1 for Ar$^+$ ions and 0.01 for O$_2^+$ ions. These values resulted in a good agreement between the experimental and simulation results in the wide parameter regime covered in this study.

The density of the $\rm{O_{2}(a^{1}\Delta_{g})}$ molecules is determined self-consistently in the simulations, based on their balance of creation, transport and de-excition at the surfaces (see details in \cite{Derzsi_2022_NeO2}). The value of the surface quenching probability is set to $\alpha = 8 \cdot 10^{-4}$ \cite{Derzsi_2022_NeO2}.

PIC/MCC simulations have been performed for the whole parameter regime covered by the PROES measurements: for driving frequencies varied between 2~MHz and 15~MHz at a peak-to-peak voltage of 350~V, for a discharge gap of 2.5~cm and a pressure of 120~Pa of the background gas, which is a mixture of 70\%~Ar and 30\%~O$_2$. The numerical parameters of the simulations were set to ensure the fulfillment of the usual PIC/MCC stability and accuracy requirements \cite{vass2022revisiting}.

\subsection{The Boltzmann term method} \label{sec:EPowerAbsorption}
For the investigation of the electron power absorption dynamics, the Boltzmann term method is applied \cite{Schulze2018_Boltzmann,lafleur2014electron,surendra1993moment}. This method provides a self-consistent, spatio-temporally resolved description of the electron power absorption by splitting the electric field into various, physically distinct terms based on the momentum balance equation, according to \cite{Vass_2020,Vass_2021}:

\begin{align}\label{Eterm}
    E_{\rm tot}&=E_{\rm in}+E_{\nabla p}+E_{\rm Ohm}, {\rm where} \nonumber \\
    E_{\rm in}&=-\frac{m_{\rm e}}{n_{\rm e}e}\left[\frac{\partial}{\partial t}(n_{\rm e}u_{\rm e})+\frac{\partial}{\partial x}(n_{\rm e}u_{\rm e}^2)\right],\nonumber \\
    E_{ \nabla p}&= - \frac{1}{n_{\rm e}e} \frac{\partial}{\partial x} p_{\parallel}, \nonumber \\	
    E_{\rm Ohm}&=-\frac{\Pi_{\rm c}}{n_{\rm e}e},
\end{align} 
with $m_{\rm e}$ and $e$ being the electron mass and the elementary charge, respectively, $n_{\rm e}$ the electron density, $u_{\rm e}$ the mean velocity, $\Pi_{\rm c}$ the electron momentum loss (as a result of collisions between electrons and the background gas particles), and $p_{\parallel}$ the diagonal element of the pressure tensor.

$E_{\rm in}$ is the electric field term originating from inertial effects, $E_{\rm Ohm}$, the Ohmic electric field, is a consequence of electron collisions, while $E_{\nabla p}$ describes (kinetic) pressure effects. This electric field term can be split into two additional terms: 
\begin{align}\label{Egradp}
E_{\nabla p}&=E_{\nabla n}+E_{\nabla T}, {\rm where} \nonumber \\
E_{\nabla n}&=-\frac{T_{\parallel}}{n_{\rm e}e}\frac{\partial n_{\rm e}}{\partial x}, \nonumber \\
E_{\nabla T}&=-\frac{1}{e}\frac{\partial T_{\parallel}}{\partial x}.
\end{align} 
Here $T_\parallel = p_\parallel/n_{\rm e}$ is the parallel electron temperature (where ``parallel'' refers to the direction of the electric field that is perpendicular to the electrode surface. The corresponding power absorption terms can be calculated based on these electric field terms by multiplying each of them with the electron conduction current density, $j_{\rm c}$.

\section{Results and discussion}\label{sec:Results}

\begin{figure}[ht]
	\centering
	\includegraphics[width=1.0\linewidth]{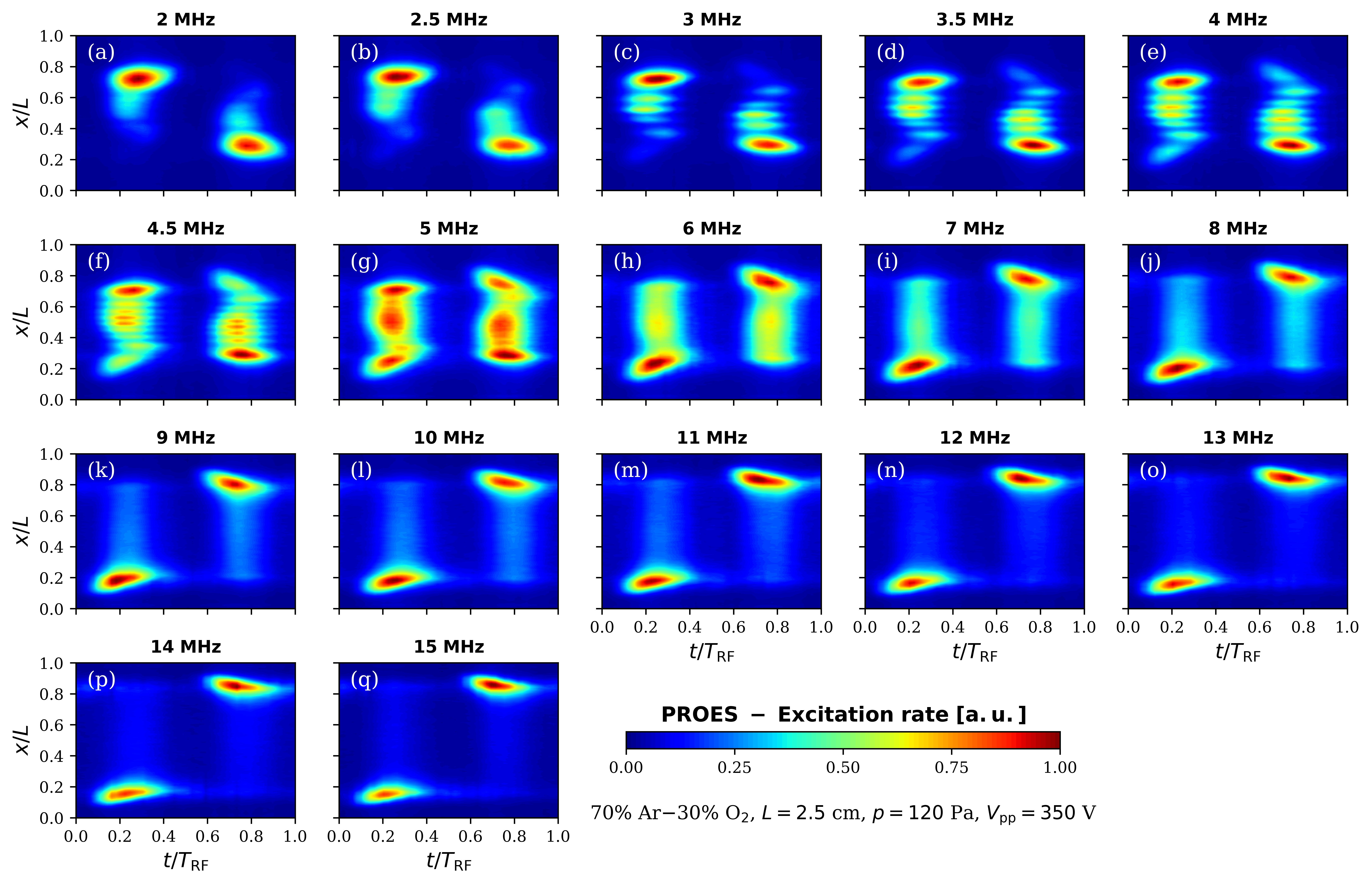}
	\caption{Spatio-temporal plots of the electron impact excitation rate from the ground state into the Ar~$\rm{2p_1}$ state measured by PROES [a.u.] for different driving frequencies (2~MHz~$\le f \le$~15~MHz), for a 70\%~Ar--30\%~O$_2$ background gas mixture. The horizontal axes correspond to one RF period, $T_{\rm{RF}} = 1/f$. The vertical axes show the distance from the powered electrode, which is located at $x/L=0$, while the grounded electrode is at $x/L=1$. The color scales of the plots are individually normalized to a maximum of 1. Discharge conditions: $L=2.5$~cm, $p=120$~Pa, $V_{\rm{pp}} = 350$~V.}
	\label{fig:exp1}
\end{figure}

Figure~\ref{fig:exp1} shows the spatio-temporal distribution of the electron impact excitation rate from the ground state into the Ar~$\rm{2p_1}$ state measured by PROES in CCPs operated
in a 70\%~Ar--30\%~O$_2$ gas mixture at different driving frequencies, $f$, between 2~MHz and 15~MHz. In all panels of figure~\ref{fig:exp1}, the vertical axes show the distance from the powered electrode, and the horizontal axes cover one RF period ($T_{\rm{RF}} = 1/f$).
At the lowest frequency of 2~MHz (figure~\ref{fig:exp1}(a)), strong excitation at the bulk side of the collapsing sheath edge is found at both electrodes, as well as significant excitation in the central bulk region, indicating discharge operation in the DA-mode. 
By increasing the frequency to 2.5~MHz, the spatio-temporal distribution of the excitation rate exhibits a spatially weakly modulated structure (figure~\ref{fig:exp1}(b)). 
As the driving frequency is further increased, these excitation structures become well separated and the gap between the excitation rate maxima decreases (the number of striations increases). 
At frequencies between 3~MHz and ~4.5~MHz (figure~\ref{fig:exp1}(c)--(f)), the relative intensity of the excitation in the bulk region (in the striated excitation patterns) is enhanced by increasing the frequency. At these driving frequencies, excitation at the expanding sheath edge can also be observed ($\alpha$-peak), as well as the formation of an excitation peak at the bulk side of the expanding sheath edge. These excitation features are also enhanced as the driving frequency is increased (figure~\ref{fig:exp1}(c)--(f)).
At 5~MHz (figure~\ref{fig:exp1}(g)), strong excitation at both the expanding and the collapsing sheath edges are found, as well as in the central bulk region. At this frequency, the gaps between the excitation rate maxima in the bulk becomes narrow, and the striations cannot be clearly resolved spatially. 
At frequencies higher than 5~MHz (6~MHz~$\le f \le$~15~MHz), the striations in the measured spatio-temporal excitation rate vanish completely in the bulk (figure~\ref{fig:exp1}(h)--(q)). The excitation at the expanding sheath edge (the $\alpha$-peak) is enhanced, while the excitation at the collapsing sheath edge and that in the bulk region get weaker by increasing the frequency, exhibiting a discharge operation mode transition from a hybrid $\alpha$-DA mode to $\alpha$-mode.\\

\begin{figure}[ht]
	\centering
 \includegraphics[width=0.8\linewidth]{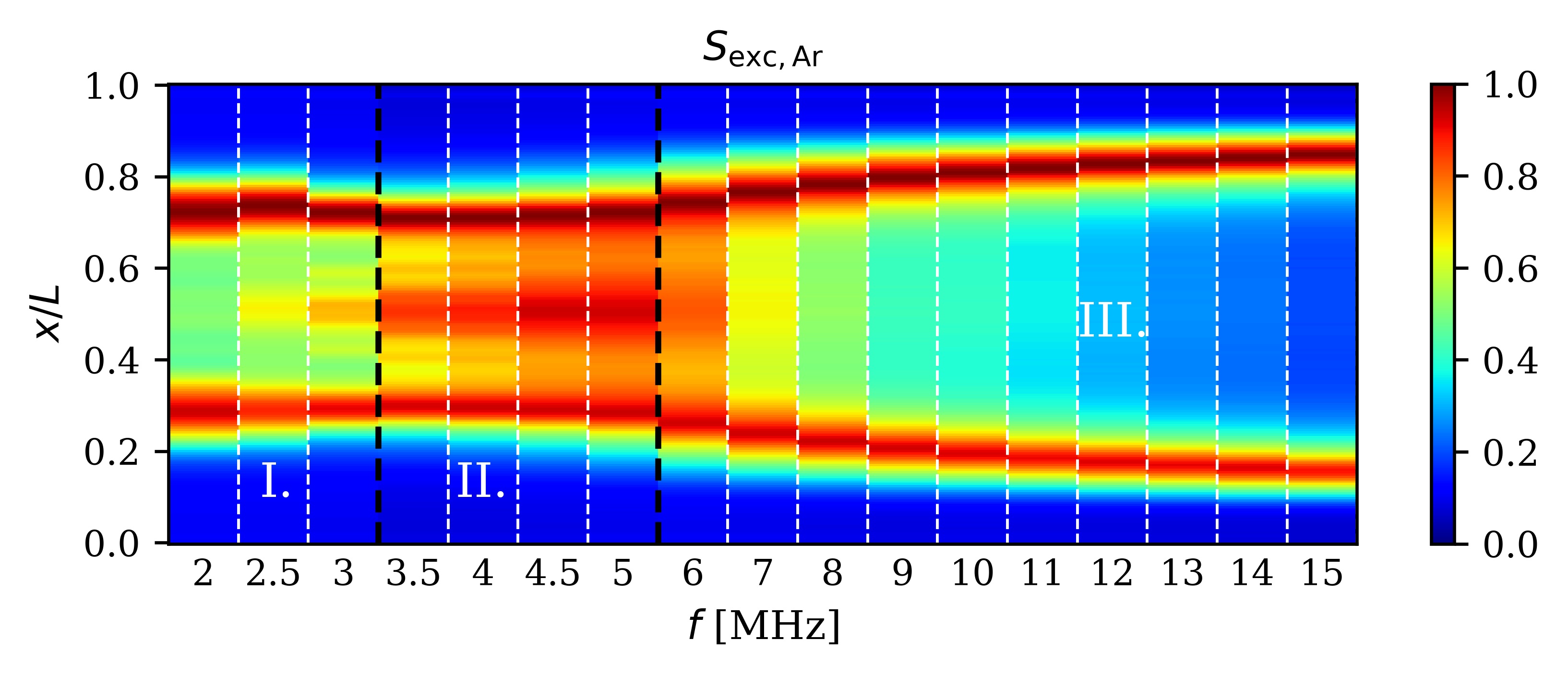}
	\caption{Time averaged results for the electron impact excitation rate from the ground state into the Ar~$\rm{2p_1}$ state measured by PROES [a.u.] for different driving frequencies (2~MHz~$\le f \le$~15~MHz), corresponding to the spatio-temporal results shown in panels of figure~\ref{fig:exp1}. The horizontal axis shows the driving frequency. The vertical axis shows the distance from the powered electrode. The results obtained for the different frequencies are separated by thin white vertical dashed lines. The thick vertical dashed black lines indicate the frequency values around which significant changes in the spatial distribution of the excitation rate take place (suggesting transitions of the discharge operation mode), defining three frequency regimes labeled as I., II. and III. for ranges 2~MHz~$\le f \le$~3~MHz, 3.5~MHz~$\le f \le$~5~MHz, and 6~MHz~$\le f \le$~15~MHz, respectively. Discharge conditions: 70\%~Ar--30\%~O$_2$ background gas mixture, $L=2.5$~cm, $p=120$~Pa, $V_{\rm{pp}} = 350$~V.}
	\label{fig:exp2}
\end{figure}

In figure~\ref{fig:exp2}, time-averaged results for the electron impact excitation rate from the ground state into the Ar~$\rm{2p_1}$ state obtained by PROES for different driving frequencies (2~MHz~$\le f \le$~15~MHz) are presented, corresponding to the spatially and temporally resolved experimental results shown in figure~\ref{fig:exp1}. The vertical axis shows the distance from the powered electrode, while the frequency values are shown in the horizontal axis. Here, columns of the time-averaged results obtained for the different frequencies are put next to each other, separated by thin white vertical dashed lines.
This figure facilitates comparison of the locations of the main excitation features revealed by PROES at different driving frequencies, as well as the analysis of the transitions between the different discharge operation modes as a function of frequency. At all frequencies, the excitation rate peaks at the sheath-bulk boundary at both electrodes. However (as discussed later), different mechanisms are responsible for these excitation maxima at low and high frequencies. 
Based on this figure, three frequency ranges can be defined with different characteristic excitation features. These are separated by thick vertical dashed black lines and labeled as I., II. and III. in the figure. Between 2~MHz and 3~MHz, the excitation is strong at the sheaths and weak (but enhanced with the frequency) in the middle of the bulk region. The 2~MHz~$\le f \le$~3~MHz frequency range is defined as range I. or low frequency range. At all frequencies in range I., the intensity of the excitation in the bulk is significantly lower than those observed at the sheath edges. As the frequency is increased from about 3.5~MHz up to about 5~MHz, the central bulk region exhibits an intensifying excitation rate/light emission, including development of striations. At these frequencies, the intensity of the excitation in the bulk is comparable to those at the sheath edges. The 3.5~MHz~$\le f \le$~6~MHz frequency range is defined as range II. or intermediate frequency range. At higher frequencies ($f\ge$~6~MHz), the excitation in the bulk decreases gradually  with increasing frequency. The 6~MHz~$\le f \le$~15~MHz frequency range is referred as frequency range III. or high frequency range. 
Figure~\ref{fig:exp2} illustrates spectacularly the effect of the driving frequency on the length of the sheath/bulk as well. The sheath length slightly increases with the frequency up to about 3.5~MHz, and decreases as the frequency is further increased.  
\\

\begin{figure}[ht]
	\centering
	\includegraphics[width=1.0\linewidth]{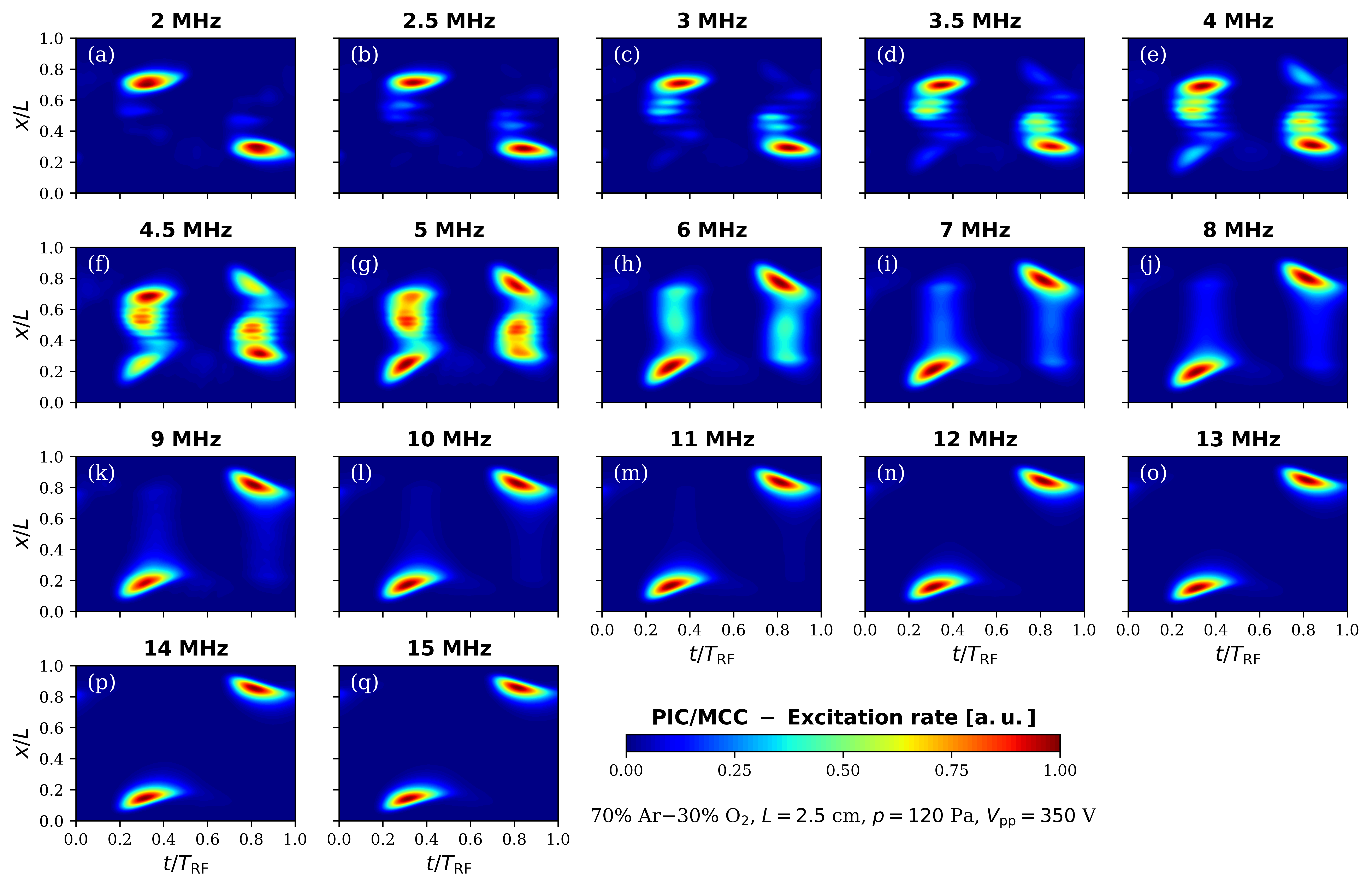}
	\caption{Spatio-temporal plots of the electron-impact excitation rate
	[a.u.] from the ground state into the Ar~$\rm{2p_1}$ state obtained from PIC/MCC simulations for different driving frequencies (2~MHz~$\le f \le$~15~MHz), for a 70\%~Ar--30\%~O$_2$ background gas mixture. The horizontal axes correspond to one RF period. $T_{\rm{RF}} = 1/f$. The vertical axes show the distance from the powered electrode, which is located at $x/L=0$, while the grounded electrode is at $x/L=1$. The color scales of the plots are individually normalized to a maximum of 1. Discharge conditions: $L=2.5$~cm, $p=120$~Pa, $V_{\rm{pp}} = 350$~V.}
	\label{fig:sim_all_f}
\end{figure}

\begin{figure}[ht]
	\centering
	\includegraphics[width=0.95\linewidth]{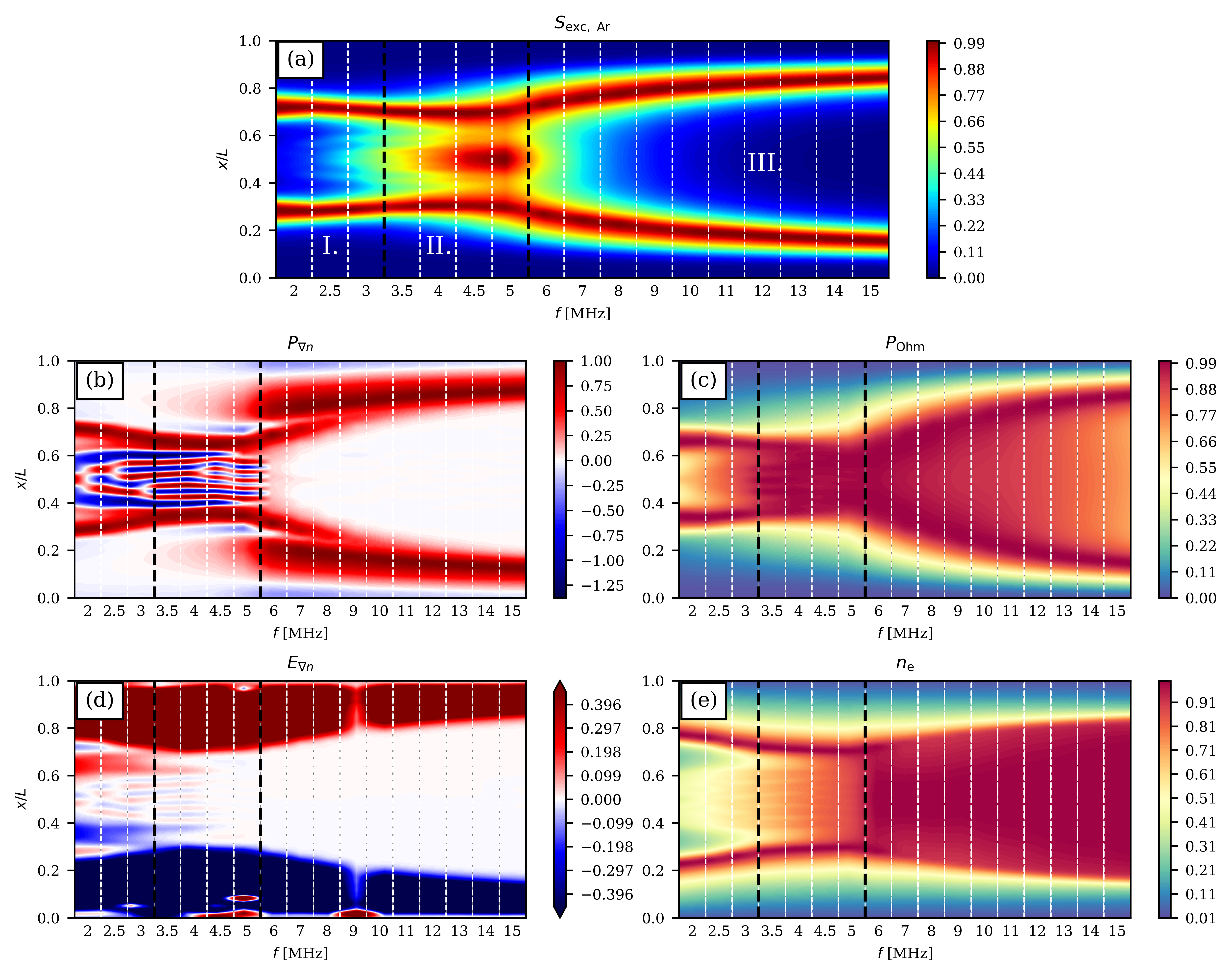}
    \caption{Time averaged results for the electron impact excitation rate from the ground state into the Ar~$\rm{2p_1}$ state, $S_{\rm exc, Ar}$ (a), the ambipolar power absorption, $P_{\nabla n}$ (b), the Ohmic power absorption, $P_{\rm Ohm}$ (c), the ambipolar electric field, $E_{\nabla n}$ (d), and the electron density, $n_{\rm e}$ (e), obtained from PIC/MCC simulations for different driving frequencies (2~MHz~$\le f \le$~15~MHz). The horizontal axes show the driving frequencies. The vertical axes show the distance from the powered electrode. At each frequency value, the corresponding quantities are normalized to their respective maxima. Discharge conditions: 70\%~Ar--30\%~O$_2$ background gas mixture, $L=2.5$~cm, $p=120$~Pa, $V_{\rm{pp}} = 350$~V. The vertical dashed black lines in all panels indicate the frequency values around which transitions in the discharge operation mode appear to take place based on the excitation rates shown in panel (a).}
	\label{fig:sim_trans}
\end{figure}

In the following, the results of PIC/MCC simulations performed for discharge conditions identical to those of the PROES measurements (see figure~\ref{fig:exp1} and figure~\ref{fig:exp2}) are presented. In figure~\ref{fig:sim_all_f}, the spatio-temporal plots of the calculated electron impact excitation rate from the ground state into the Ar~$\rm{2p_1}$ state are shown for different driving frequencies (2~MHz~$\le f \le$~15~MHz). At all frequencies, the main excitation patterns revealed by PROES are well reproduced by the simulations. 
At low frequencies, in agreement with the PROES measurements, the excitation rate in the bulk region is spatially modulated. The variation of the number of striations as a function of the driving frequency is also well reproduced by the simulations. At 2.5~MHz, besides the strong excitation at the bulk side of the collapsing sheath edge at both electrodes, 2 weak excitation rate maxima can be observed in the central bulk region during each half of the RF period. The number of striations increases with the frequency, exhibiting 5 peaks in the excitation rate in the central bulk region at 4.5~MHz, in perfect agreement with the experimental results. The enhancement of the excitation at the expanding sheath edge, as well as the formation and enhancement of an excitation peak at the bulk side of the expanding sheath edge with increasing frequency are also captured in the simulations. Conforming the PROES results, at frequencies above 5~MHz, the excitation in the bulk region is reduced and strong excitation is found at the expanding sheath edges. 

In figure~\ref{fig:sim_trans}, time averaged results for several discharge characteristics obtained from PIC/MCC simulations for different driving frequencies (2~MHz~$\le f \le$~15~MHz) are presented: panel (a) shows the electron impact excitation rate from the ground state into the Ar~$\rm{2p_1}$ state, $S_{\rm exc, Ar}$; the ambipolar and Ohmic power absorptions, $P_{\nabla n}$ and $P_{\rm Ohm}$, respectively, are shown in panels (b) and (c); panel (d) shows the ambipolar electric field, $E_{\nabla n}$, while in panel (e) the electron density, $n_{\rm e}$, is plotted for different driving frequencies. The format of the panels is the same as that used in case of figure~\ref{fig:exp2}. Here (unlike in figure~\ref{fig:exp2}), the results obtained for the different driving frequencies are presented by using a color map in which the intermediate values are smoothed (resulting in continuous transitions between the results corresponding to the different driving frequencies, contrary to the step-like transitions one can see in figure~\ref{fig:exp2}). Thin white vertical dashed lines are used here again to separate the slabs (columns) of time-averaged results corresponding to the different driving frequencies. 
The plot in panel (a) looks similar to the time-averaged PROES results shown in figure~\ref{fig:exp2}. The locations of the main excitation patterns in the gap observed in the experimental results at different frequencies are well reflected by the simulations. This way of representation of the simulation results on the electron impact excitation rate exposes impressively the differences between the characteristic excitation features at distinct frequencies and reveals the frequencies (frequency regimes) where significant changes in the spatial distribution of the excitation rate take place. Such pronounced changes in the excitation or the ionization rates (judged by simply looking at the spatio-temporal distribution of the excitation/ionization rate) are generally associated with changes in the dominant electron power absorption mechanisms and transitions in the discharge operation mode. 
Based on the results shown in panel~(a), two transitions in the electron power absorption and excitation
happen as the frequency is varied. 
The transitions seem to occur at frequencies around 3~MHz and 5~MHz, in accordance with the PROES results (figure~\ref{fig:exp2}). These frequencies are marked by vertical black dashed lines in all panels of figure~\ref{fig:sim_trans}. Similarly to the experimental results, one can identify a frequency range where strong excitation is found at the sheath edges and no excitation/only weak excitation in the middle of the bulk ($f\le3.5$~MHz, range I.), a range where strong excitation can be observed both at the sheath edges and in the bulk region (3.5~MHz~$\le f \le$~5~MHz, range II.), and one range where strong excitation is found at the sheath edges again ($f\ge~6$~MHz, range III.). 

In panel (b), which shows the ambipolar power absorption term, striations can be observed in the bulk region at low frequencies (range I.). As the frequency is increased, the striations branch to other striations (the number of striations increases up to about 5~MHz, i.e. striations are present in both frequency ranges I. and II.). These bifurcations arise here from the specific way of representation of the simulation data using interpolation, however finer resolution of the low frequency range studied here would also reveal these structures. At high frequencies (range III.), the ambipolar power absorption is concentrated near the sheaths, characteristic of the $\alpha$-mode.
The Ohmic power absorption (panel~(c)) is relatively high in the bulk in frequency range II. (characteristic of the DA-mode, resulting in strong excitation in the central bulk region) and decreases in the middle of the bulk by decreasing/increasing the driving frequency in range I./range III. (the excitation in the bulk region disappears by decreasing/increasing the frequency towards range I./III.). The Ohmic power absorption is high whenever the electron density decreases locally (see the time-averaged electron densities in panel~(e)). By increasing the frequency (in ranges I. and II.), more and more striations develop (at positions of local density minima). The density decrease is sharper at lower number of striations (at lower frequencies), where local maxima in the Ohmic power absorption are also resolved (at $f=3.5$~MHz in panel~(c)). 
Whenever the number of striations is high, the density decrease will not be as large as in case of lower number of striations, therefore striations in the Ohmic power absorption are not visible in the intermediate frequency range II. for $f\ge4$~MHz. The change in the number of striations is also the reason why there is a region of negative/positive ambipolar electric field, $E_{\nabla n}$, near the powered/grounded electrode, respectively, as seen in panel~(d): since the smaller number of striations leads to a steeper increase/decrease of electron density, the corresponding density gradients will be high, which leads to an increase in the magnitude of the ambipolar electric field. Based on these findings, there is no change in the dominant power absorption mechanisms in frequency range I. and II. (at the two sides of the black vertical dashed line at 3~MHz), despite the significant differences that can clearly be observed in the excitation rates. 
Based on the excitation rate (panel~(a)) one could identify mode transitions from $\alpha$-mode to the DA-mode and striation mode as the frequency is decreased from 15~MHz to 2~MHz. However, regarding the power absorption, it is not straightforward to infer the power absorption mode transitions based on the excitation rate alone. The present results clearly show that the same electron power absorption mechanisms could be associated with excitation patterns of significantly different characteristics.
\\

\begin{figure}[ht]
	\centering
	\includegraphics[width=0.95\linewidth]{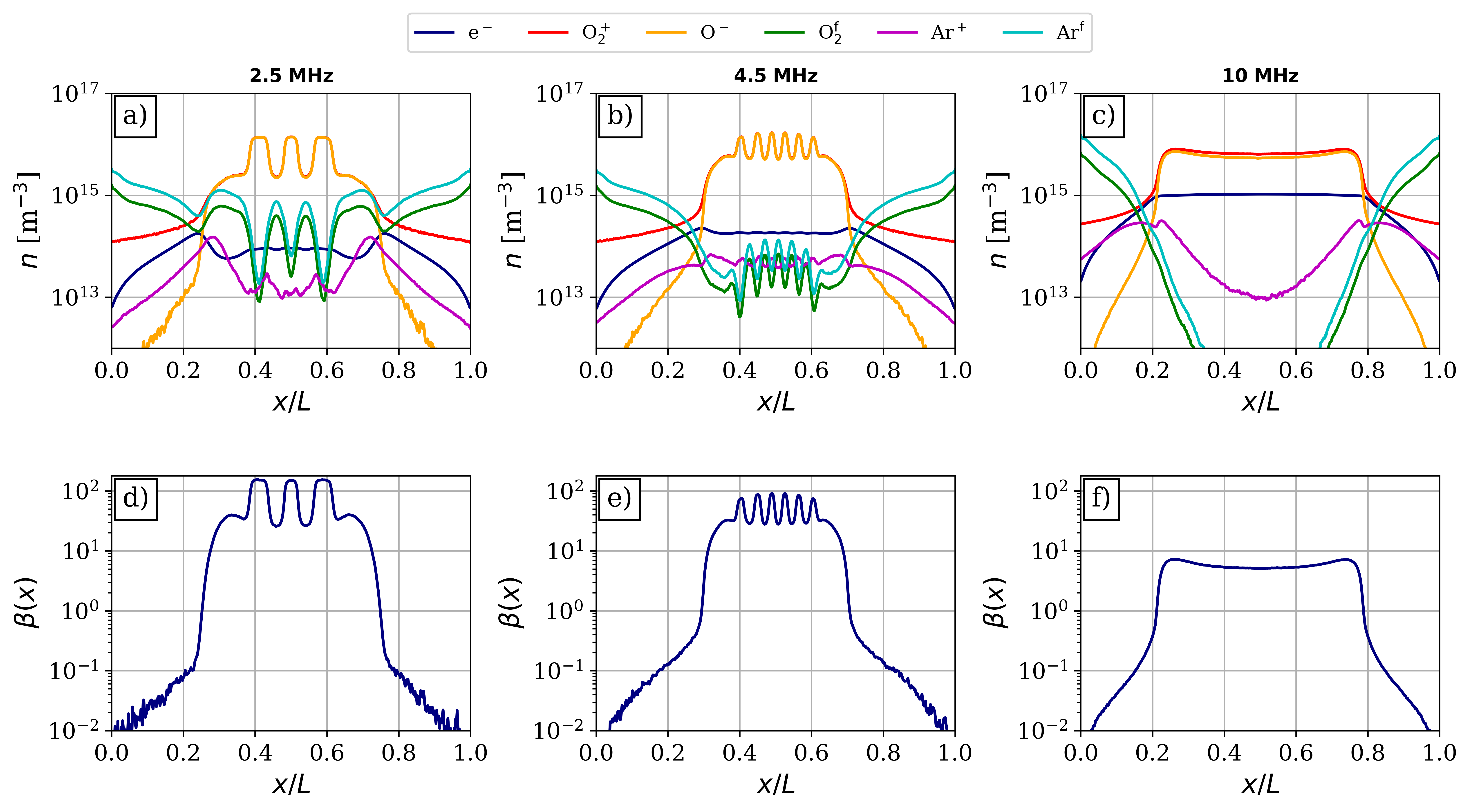}
    \caption{Time averaged particle density distributions obtained from PIC/MCC simulations for different driving frequencies: (a) 2~MHz, (b) 4.5~MHz, and (c) 10~MHz and ratio of the time averaged negative ion (O$^-$) density and electron density (local electronegativity, $\beta$(x)): (d) 2~MHz, (e) 4.5~MHz, and (f) 10~MHz. 
    Discharge conditions: 70\%~Ar--30\%~O$_2$ background gas mixture, $L=2.5$~cm, $p=120$~Pa, $V_{\rm{pp}} = 350$~V.}
	\label{fig:sim_dens}
\end{figure}

\begin{figure}[ht]
	\centering
	\includegraphics[width=0.95\linewidth]{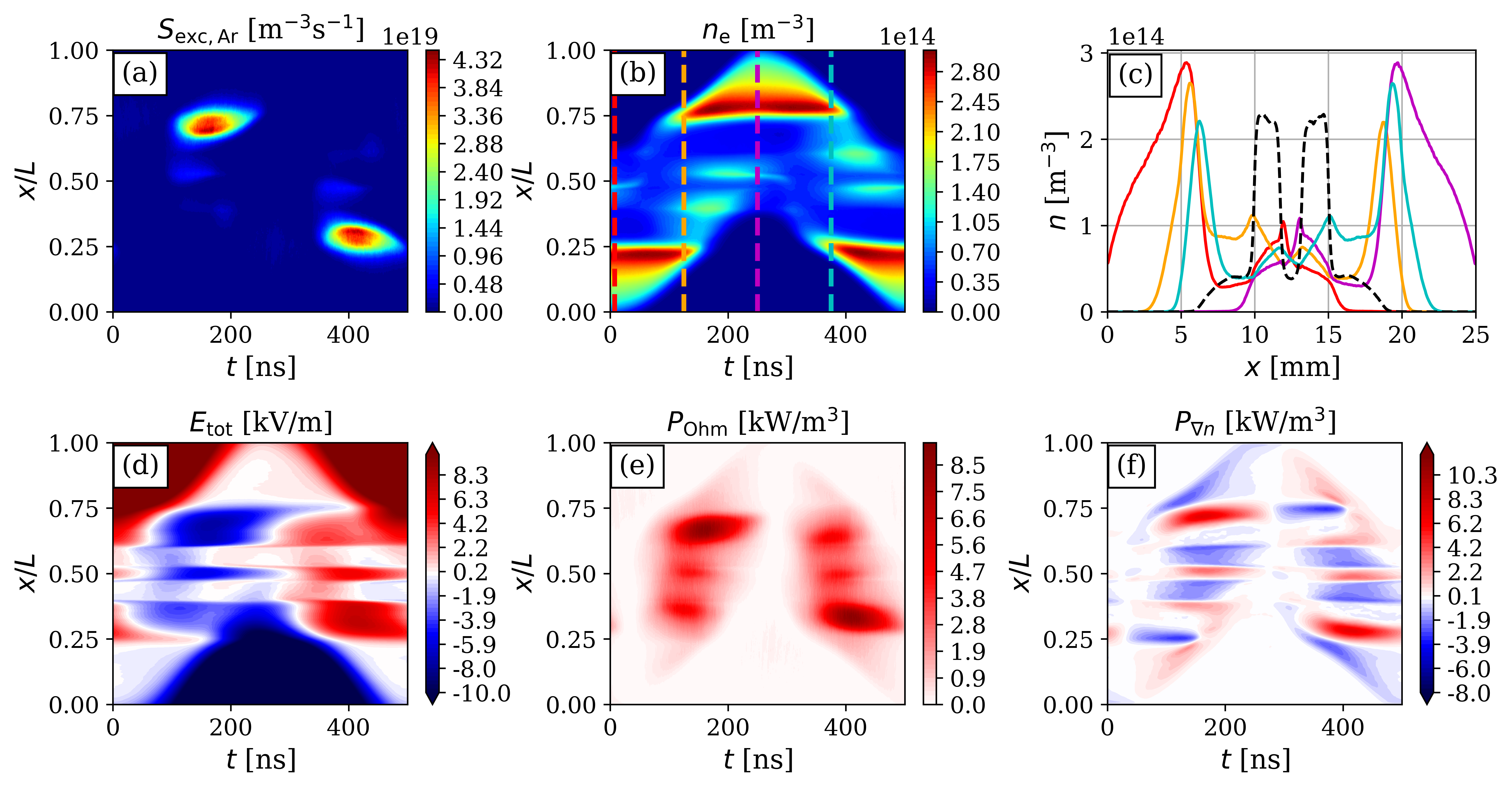}
	\caption{Spatio-temporal plots of the electron-impact excitation rate, $S_{\rm exc, Ar}$ (a), the electron density, $n_{\rm e}$ (b), temporal snapshots of the electron density (solid color curves) as well as the temporally averaged negative ion (O$^-$) density (dashed black curve) divided by a factor of 50 (c), spatio-temporal plot of the electric field, $E_{\rm tot}$ (d), the Ohmic power absorption, $P_{\rm Ohm}$ (e), and the ambipolar power absorption, $P_{\nabla n}$ (f) for a driving frequency of $f=2$~MHz . The colors of the curves in panel (c) denote the respective time instances in panel (b). The powered electrode is located at $x/L=0$, while the grounded electrode is at $x/L=1$. Discharge conditions: 70\%~Ar--30\%~O$_2$ background gas mixture, $L=2.5$~cm, $p=120$~Pa, $V_{\rm{pp}} = 350$~V.} 
	\label{fig:details_2MHz}
\end{figure}

\begin{figure}[ht]
	\centering
	\includegraphics[width=0.95\linewidth]{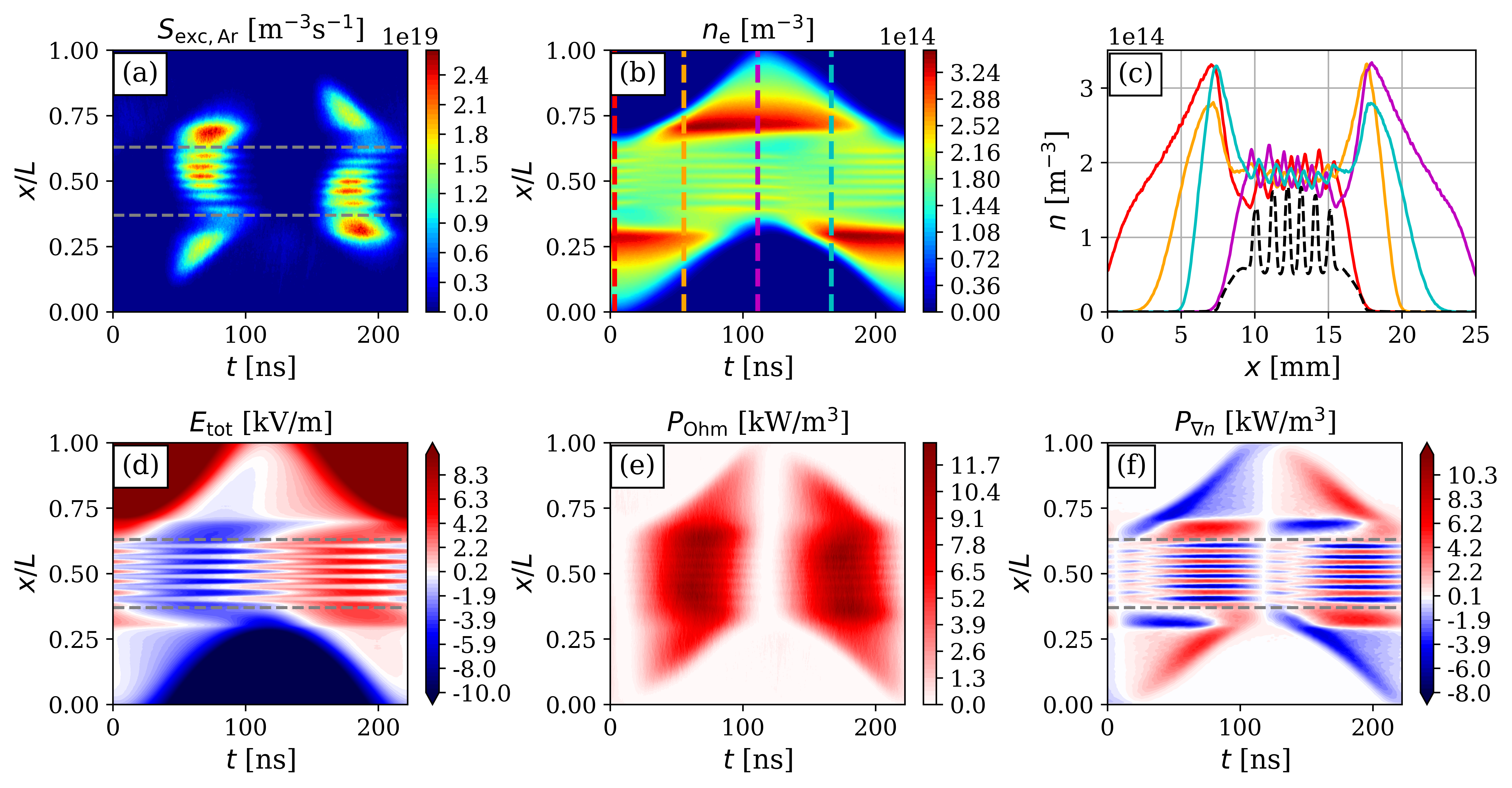}
	\caption{Various discharge characteristics -- same as those presented in figure~\ref{fig:details_2MHz} -- obtained for a driving frequency of $f=4.5$~MHz. Here, the temporally averaged negative ion (O$^-$) density (dashed black curve in panel (c)) is divided by a factor of 100. Discharge conditions: 70\%~Ar--30\%~O$_2$ background gas mixture, $L=2.5$~cm, $p=120$~Pa, $V_{\rm{pp}} = 350$~V. The horizontal dashed black lines in panels (a), (d) and (f) indicate the region where striations develop in the bulk.} 
	\label{fig:details_4.5MHz}
\end{figure}

\begin{figure}[ht]
	\centering
	\includegraphics[width=0.95\linewidth]{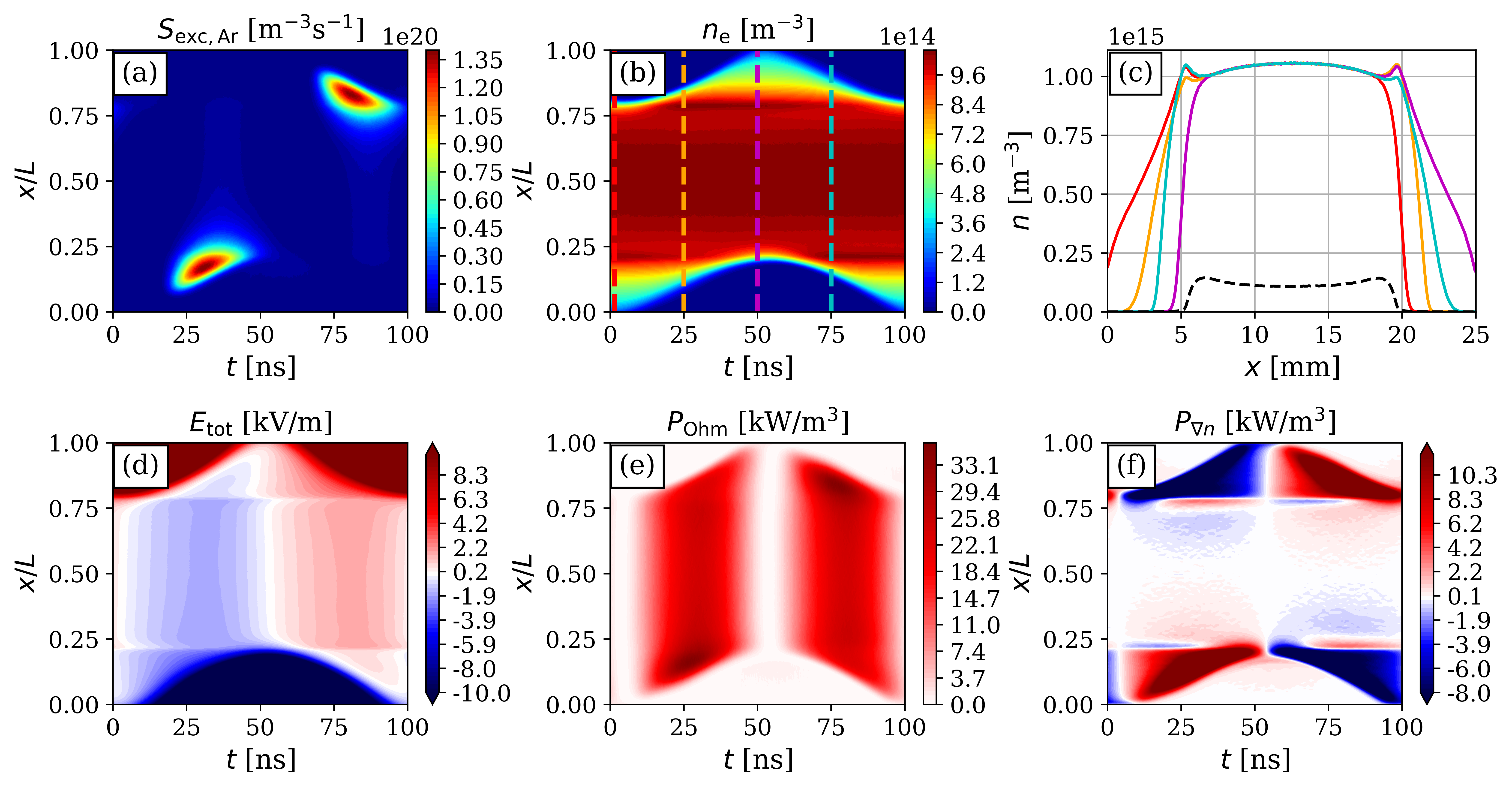}
	\caption{Various discharge characteristics -- same as those presented in figure~\ref{fig:details_2MHz} -- obtained for a driving frequency of $f=10$~MHz. Here, the temporally averaged negative ion (O$^-$) density (dashed black curve in panel (c)) is divided by a factor of 100. Discharge conditions: 70\%~Ar--30\%~O$_2$ background gas mixture, $L=2.5$~cm, $p=120$~Pa, $V_{\rm{pp}} = 350$~V.} 
	\label{fig:details_10MHz}
\end{figure}

In the following, the PIC/MCC simulation results are analysed in detail for thee different driving frequencies: (i) 2~MHz, (ii) 4.5~MHz, and (iii) 10~MHz. These frequency values are in the frequency ranges I. (low frequency range), II. (intermediate frequency range), and III. (high frequency range), respectively. At these driving frequencies, the corresponding excitation rates, obtained both from PROES (see figure~\ref{fig:exp1} and figure~\ref{fig:exp2}) and PIC/MCC simulations (see figure~\ref{fig:sim_all_f} and figure~\ref{fig:sim_trans}) exhibit substantially different characteristics, suggesting different discharge operation modes. For these three cases, the time averaged particle density distributions and the electronegativity are plotted in figure~\ref{fig:sim_dens}, while various discharge characteristics, including the spatio-temporal distribution of the electron-impact excitation rate, the electron density, the electric field, the ambipolar and Ohmic power absorption terms, and temporal snapshots of particle densities,
are shown in figures~\ref{fig:details_2MHz} (2~MHz), \ref{fig:details_4.5MHz} (4.5~MHz), and \ref{fig:details_10MHz} (10~MHz), respectively.

At 2~MHz, local maxima at the edges of the bulk region can be observed in the time averaged electron density distribution (figure~\ref{fig:sim_dens}(a)). The electron density as well as the density of heavy particles exhibit spatial modulation in the discharge center. The electron density is significantly lower than the density of negative O$^-$ ions in the bulk, resulting in high electronegativity in the bulk (figure~\ref{fig:sim_dens}(d)). Due to the spatial modulation of the particle densities in the bulk, the electronegativity also shows this feature. The ratio of the negative ion density and electron density is above 30 in the middle of the gap, reaching high values of about 160 at the positions of electron density minima in the bulk. Under these conditions, the global electronegativity of the discharge (the ratio of the density of negative ions and electrons averaged over the electrode gap) is about 30. The local minima/maxima of the fast neutrals is due to the presence of an oscillating electric field, as a result of the presence of the striations: since the charged heavy particle density has a gradient, so will the electron density, which leads to an ambipolar electric field. This field will accelerate charged particles, which can create fast neutrals through collisions with the background gas. The maxima of the fast neutral density correspond to maxima of the electric field (see figure \ref{fig:details_2MHz}). The local maxima of the Ar$^+$ density in the center of the discharge is due to electrons being accelerated by the ambipolar electric field, as will be discussed later.
At the higher frequency of 4.5~MHz (figure~\ref{fig:sim_dens}(b)), similarly to the case of 2.5~MHz, local maxima in the time averaged electron density distribution at the edges of the bulk and high negative ion density in the bulk can be seen, as well as spatial modulation of the particle densities in the discharge center. However, the electron density and the density of Ar$^+$ ions is enhanced in the discharge center at this frequency, while the density of fast neutrals decreases in the bulk, in accordance with the higher number of striations present in this situation. The electronegativity is modulated in space, with maximum values of about 90 in the center of the bulk (figure~\ref{fig:sim_dens}(e)). The global electronegativity is about 16 at this frequency. 
At 10~MHz (figure~\ref{fig:sim_dens}(c)), the bulk region is wider compared to the lower frequency cases. The time averaged electron density peaks in the discharge center, while the density of Ar$^+$ ions as well as that of fast neutrals drop in the bulk. The O$^-$ and O$_2^+$ densities are high in the bulk, the corresponding time averaged density profiles exhibit local minima in the discharge center and local maxima near the bulk-sheath boundary. As a result of this, the electronegativity exhibits local maxima of about 7 at the bulk edges and a local minimum of about 5 in the center of the bulk (figure~\ref{fig:sim_dens}(f)). The global electronegativity is about 3 at this frequency. The electronegativity shows similar spatial distribution also at higher frequencies. The global electronegativity of the discharge decreases with further increase of the frequency, and is about 1.2 at the highest frequency of 15 MHz studied here.

Figure~\ref{fig:details_2MHz} shows PIC/MCC simulation results for various discharge characteristics obtained for the 70\%~Ar--30\%~O$_2$ background gas mixture at $p=120$~Pa pressure and $f=2$~MHz driving frequency. The spatio-temporal distribution of the electron-impact excitation rate from the ground state into the Ar~$\rm{2p_1}$ state, $S_{\rm exc}$, is shown in panel (a) (not normalized, otherwise same as panel (a) of figure~\ref{fig:sim_all_f}), while that of the electron density, $n_{\rm e}$, is shown in panel (b). In panel (b), 4 time instances within the RF period (at values of $t/T_{\rm RF}$ of 0, 0.25, 0.5, and 0.75) are marked with vertical dashed lines in different colours. The electron densities in the discharge gap corresponding to these time instances are shown in panel (c), where the colors of the solid curves identify the respective time instances in panel (b). 
In panel (c) the temporally averaged density of negative O$^-$ ions divided by a factor of 50 is also included (dashed black line), to illustrate that the density of O$^-$ ions is much higher than the electron density in the bulk region. The spatio-temporal plots of the electric field, $E_{\rm tot}$, the Ohmic power absorption, $P_{\rm Ohm}$, and the ambipolar power absorption, $P_{\nabla n}$, are shown in panels (d), (e) and (f), respectively. 
At all selected $t/T_{\rm RF}$ time instances, strong electron density peaks can be observed at the edges of the bulk region: one peak at the powered/grounded electrode side at $t/T_{\rm RF}$ values of 0 and 0.5 and peaks at both sides at $t/T_{\rm RF}$ values of 0.25 and 0.75. The electron density is low in the discharge center, where additional local electron density peaks can be observed: one local peak at $t/T_{\rm RF}$ values of 0 and 0.5 and two local peaks at $t/T_{\rm RF}$ values of 0.25 and 0.75. Under these conditions, the ratio of the O$^-$ ion density and the electron density, i.e. the electronegativity is high (between about 30 and 160) in the discharge center (see figure~\ref{fig:sim_dens}(d)). Due to the high electronegativity in the discharge center, the conductivity of the plasma is low in the bulk. As a consequence of this, the Ohmic power absorption, $P_{\rm Ohm}$, is high in the discharge center at the times of electron density minima within the RF period (panel~(e)). The ambipolar power absorption, $P_{\nabla n}$, peaks at the edges of the bulk region at the time of sheath collapse at both electrodes (panel (f)) due to the local maxima (panels (b) and (c)).
These ambipolar power absorption maxima near the sheath edges produce an electric field which does not change sign during the RF-cycle (e.g., near the powered electrode, the ambipolar electric field is always negative). This is a well-known feature of eletronegative discharges \cite{Vass_2020}, and is due to the fact that the ``electropositive edge'', i.e. the local electron density maximum near the sheath is not modulated in time (cf. also panel (b)). It is very important to note that the ambipolar electric field near the striations are of different nature. Since, based on panel (b) and (c), the electron density is temporally modulated, so is the density gradient. Thus, at a given position in the bulk, the ambipolar electric field will change sign depending on the sign of the electron conduction current.
This means that in the bulk region, the ambipolar electron power absorption has the same sign in the whole RF-cycle at a given position. This difference between the behaviour of the ambipolar power absorption near the sheath edge and in the bulk is what leads to the spatio-temporal excitation patterns seen in panel (a): the maxima near the sheath edges are a consequence of electrons accelerated by the local ambipolar electric field, as a result of the density gradient due to the presence of the ``electropositive edge'' (cf. panel (f)). 
The second peak near $x/L=0.5$ is due to the ambipolar electric field resulting from the presence of the striations: since the electron power absorption at this position is positive in the entire RF-cycle, the local excitation maxima are much closer to each other: the one in the first half of the RF-cycle is closer to the grounded electrode, since electrons move towards the grounded electrode, while the situation is the opposite in the second half of the RF-cycle. This fact is the reason for the symmetric local maxima in the Ar$^+$-density seen in fig. \ref{fig:sim_dens} (a). Under the present conditions, both the ambipolar and the Ohmic power absorption mechanisms contribute to the power absorption, their magnitudes being similar. 

Similarly to the 2~MHz case (figure~\ref{fig:details_2MHz}), some discharge characteristics obtained for 4.5~MHz are shown in figure~\ref{fig:details_4.5MHz}. In this case, the spatio-temporal distribution of the excitation rate shows strong excitation at the collapsing sheath edge, spatially modulated structure (with 6 excitation peaks) in the central bulk region and excitation at the expanding sheath edge as well (panel (a)). The electron density profiles plotted at different time instances (panel (c)) show strong peaks at the edges of the bulk region (one peak at the powered/grounded electrode side at $t/T_{\rm RF}$ values of 0 and 0.5 and peaks at both sides at $t/T_{\rm RF}$ values of 0.25 and 0.75). The electron density is low and spatially modulated in the bulk (panels (b) and (c)). This is similar to the 2~MHz case. However, the electron density is enhanced in the discharge center and the number of the local electron density peaks is higher compared to the results obtained for 2~MHz: here, 6 local electron density peaks can be observed at all selected $t/T_{\rm RF}$ time instances, which is due to the increased driving frequency and the correspondingly smaller amplitude of ion charge separation, which causes the striations \cite{Liu2016}. The ratio of the O$^-$ ion density and the electron density is high in the discharge center (see figure~\ref{fig:sim_dens}), exhibiting spatial modulation with values above 30 in the bulk and reaching maximum values of about 90 in the center of the bulk.
Similarly to the previous case, the Ohmic power absorption is high in the discharge center (panel (e)) and shows spatial modulation in the bulk. The ambipolar power absorption peaks at the edges of the bulk region and exhibits striations in the bulk (panel (f)). The six peaks in panel (a) in the bulk are the result of striations, as in the previous case: the local maxima are due to the positive ambipolar power absorption near each striation as a result of the electron density gradient. The superposition of the Ohmic and ambipolar electric fields results in a strong electric field at both sides of the bulk as well as in the bulk region (panel (d)).

For 10~MHz, the simulation results on various discharge characteristics are shown in figure~\ref{fig:details_10MHz}. At this frequency, strong excitation is found only at the expanding sheath edges (panel~(a)). The electron density profiles for the different time instances show only small local minima at the edges of the bulk region (panel~(c)). The time averaged O$^-$ density has a local minimum in the center and local maxima at the sheath-bulk boundary. The ratio of the O$^-$ ion density and the electron density is low in the discharge center (see figure~\ref{fig:sim_dens}(f)), with local minimum of about 5.
The Ohmic power absorption (panel~(e)) shows peaks at the expanding sheath edges. The ambipolar power absorption (panel (f)) is also concentrated at the sheath edges at both electrodes, which is characteristic of the $\alpha$-mode. Since in this case there are no striations, and due to the low electronegativity the ``electropositive edge'' is not pronounced, the only peak visible in panel~(a) is the $\alpha$-peak, although there is a small spatio-temporal region of ambipolar power absorption according to panel (f) near the collapsing phase of each sheath. However, this is too small to lead to any excitation that would be visible in panel~(a).

\section{Conclusions}\label{sec:Conclusions}
Phase Resolved Optical Emission Spectroscopy (PROES) measurements combined with 1d3v Particle-in-Cell/Monte Carlo Collisions (PIC/MCC) simulations have been performed in low-pressure capacitively coupled argon-oxygen plasmas. The discharge conditions covered a wide frequency range between 2~MHz and 15~MHz in a geometrically symmetric plasma reactor with a gap length of 2.5~cm, operated in a mixture of 70\%~Ar and 30\%~O$_2$ (volumetric ratio) at 120~Pa, applying a peak-to-peak voltage of 350~V. 

The measured electron impact excitation rates from the Ar ground state into the Ar~$\rm{2p_1}$ state were compared to the PIC/MCC simulation results on the Ar excitation rate, showing a good qualitative agreement for all discharge conditions.
At the lowest frequency of 2~MHz, strong excitation at the bulk side of the collapsing sheath edge was found at both electrodes, as well as weak excitation in the
central bulk region, indicating discharge operation in the DA-mode. By increasing the
frequency, the spatio-temporal distribution of the excitation rate was found to exhibit
spatially modulated excitation patterns, with increasing number of these features with the frequency, as well as enhancement of the excitation in the bulk and at the expanding sheath edges. At frequencies higher than 5~MHz, it was found that the striations vanish in the bulk, while the excitation at the expanding sheath edge is enhanced, and the excitation in the bulk and at the collapsing sheath edge becomes weaker. 

Visualization of the time-averaged results for the electron impact excitation rate from the Ar ground state into the Ar~$\rm{2p_1}$ state (obtained both by PROES and PIC/MCC simulations) for different driving frequencies in a single plot clearly showed the main differences in the characteristic excitation features at different frequencies and it revealed the frequency values around which significant changes in the excitation rate take place. Such changes are generally considered to be predictive of transitions of the dominant electron power absorption mode and discharge operation mode. 
Three frequency ranges could be defined with profoundly different characteristic excitation features. Up to 3~MHz (frequency range I., low frequency range), the excitation was found to be strong at the sheaths and weak in the middle of the bulk region. At frequencies between 3.5~MHz and about 5~MHz (frequency range II., intermediate frequency range), the excitation was found to be strong at the sheaths, while the central bulk region was found to exhibit intensifying excitation rate including the formation of spatially modulated patterns. At frequencies higher than 5~MHz (frequency range III., high frequency range), the excitation in the bulk was found to decrease gradually, while the excitation was found to remain strong at the sheath edges. This suggested two transitions in the dominant power absorption mode and the discharge operation mode as the frequency was tuned between 2~MHz and 15~Mhz: one transition at about 3~MHz and another one at about 5~MHz. 

Based on Boltzmann term analysis, the mechanisms behind the excitation characteristics at different frequencies were analyzed. 
At low frequencies (range I.), striations could be observed in the ambipolar power absorption in the bulk region. As the frequency was increased, the striations were found to branch to other striations at intermediate frequencies (range II.). At high frequencies (range III.), the ambipolar power absorption was found to be concentrated near the sheaths.  The Ohmic power absorption was found to be high in the bulk in frequency range II. and to decrease in this region by decreasing/increasing the driving frequency towards range I./range III. The variation of the number of striations with the frequency was found to have important effects on the contributions of the Ohmic and ambipolar terms to the power absorption. 
It was found that despite the significantly different excitation maps seen in the different frequency regimes, the dominant power absorption mechanisms are basically the same in frequency ranges I. and II. The present results clearly showed that it is not straightforward to infer the power absorption mode transitions based on the excitation rate alone. It was found that the same electron power absorption mechanisms could be associated with excitation patterns of significantly different characteristics.

The agreement between the experimental data and the results of the simulations over a wide range of operation frequencies confirms that the discharge model properly captures the main physical phenomena in the CCP operated in the Ar-O$_2$ mixture and verifies the implementation of this model in the simulation code.

\ack The authors thank Ihor Korolov for his invaluable advice on the cross sections used in the simulations. This work was supported by the Hungarian National Research, Development and Innovation Office via grants K-134462 and FK-128924, by the German Research Foundation (DFG) within the frame of the collaborative research centre SFB-CRC 1316 (project A4) and by the \'{U}NKP-22-3 New National Excellence Program of the Ministry for Innovation and Technology from the source of the National Research, Development and Innovation Fund. 

\section*{References}

\bibliography{references}
\bibliographystyle{iopart-num}

\end{document}